\documentclass[11pt]{article}
\usepackage{a4}
\usepackage{amsmath}
\usepackage{amssymb}
\usepackage[dvips]{graphicx}
\usepackage{subfigure}
\def\beq{\begin{equation}}
\def\eeq{\end{equation}}
\def\beqa{\begin{eqnarray}}
\def\eeqa{\end{eqnarray}}
\def\n{\nonumber \\}

\newcommand {\tr}{{\rm tr}\,}
\newcommand {\Tr}{\mbox{Tr}\,}
\newcommand {\CTr}{{\cal T}r\,}

\def\dag{\dagger}
\newcommand{\id}{{1\!\!1}} 

\begin{document}

\vspace*{1.0cm}
\begin{flushright}
{SAGA-HE-262}
\end{flushright}
\vskip 1.0cm

\begin{center}
{\large{\bf Ginsparg-Wilson relation on a fuzzy 2-sphere for adjoint matter}}
\vskip 1.0cm

{\large Hajime Aoki\footnote{e-mail
 address: haoki@cc.saga-u.ac.jp}
}
\vskip 0.5cm

{\it Department of Physics, Saga University, Saga 840-8502,
Japan  }\\

\end{center}

\vskip 1cm
\begin{center}
\begin{bf}
Abstract
\end{bf}
\end{center}
We formulate a Ginsparg-Wilson relation on a fuzzy 2-sphere for
matter in the adjoint representation of the gauge group.
Because of the Ginsparg-Wilson relation, an index theorem is satisfied.
Our formulation is applicable to topologically 
nontrivial configurations as monopoles.
It gives a solid basis for obtaining
chiral fermions,
which are an important ingredient of the standard model, from
matrix model formulations of the superstring theory,
such as the IIB matrix model,
by considering topological configurations in the extra dimensions.
We finally discuss whether this mechanism really works.

\newpage
\setcounter{footnote}{0}
\section{Introduction}
\setcounter{equation}{0}

Matrix models are a promising candidate to 
formulate the superstring theory nonperturbatively \cite{Banks:1996vh,IKKT},
and they indeed include quantum gravity and gauge theory.
One of the important subjects in such studies is to 
connect these models to phenomenology. 
Spacetime structures can be analyzed dynamically 
in the IIB matrix model \cite{Aoki:1998vn},
and four dimensionality seems to be preferred \cite{Aoki:1998vn,Nishimura:2001sx}.
Assuming four-dimensional spacetime is obtained, 
we next want to show the standard model of
particle physics on it.
A crucial issue for it is to realize chiral fermions,
which also ensures the existence of massless fermions.
Without chiral symmetries, quantum corrections would induce mass
of order of the Planck scale in general.

A way to obtain chiral spectrum in our spacetime 
is to consider topologically nontrivial configurations 
in the extra dimensions\footnote{Having
this mechanism in mind, we analyzed dynamics of a model on a fuzzy 2-sphere
and showed that topologically nontrivial configurations are indeed realized  \cite{AIMN}.
Models of four-dimensional field theory with fuzzy extra dimensions were
studied in
\cite{Steinacker:2007ay,Chatzistavrakidis:2009ix}.}.
Owing to the index theorem \cite{Atiyah:1971rm},
topological charge of the background 
provides the index of the Dirac operator,
{\it i.e.}, 
the difference of the numbers of chiral zero modes,
which then produce massless chiral fermions in our spacetime.
Generalizations of the index theorem 
to matrix models or noncommutative spaces are,
however, mostly formulated
in spaces with an infinite size,
and it is widely believed that topological charges
cannot be defined in a system with finite degrees of freedom.

The situation is similar to the lattice gauge theories,
where the theory is defined on a finite number of 
lattice points.
There a problem to properly define the chiral symmetry and 
the index theorem arises
due to the doubling problem
\cite{Nielsen:1980rz}.
The problem has been solved successfully 
by introducing Dirac operators
satisfying a Ginsparg-Wilson (GW)
relation \cite{GinspargWilson}.
While all the gauge field configurations are continuously 
connected and there seems to be no room for defining 
separate topological sectors, 
the configuration space becomes disconnected by 
introducing the admissibility condition
and the various topological sectors
can then be realized
\cite{Luscher:1981zq}.

The ideas of using the GW relation were applied 
to matrix models or noncommutative geometries.
In ref. \cite{AIN2}, we have provided 
a general prescription to construct 
a GW Dirac operator with coupling to background gauge fields. 
As a concrete example,
a GW Dirac operator on a fuzzy 2-sphere \cite{Madore}
was given\footnote{
A GW Dirac operator without gauge field backgrounds
was given earlier in \cite{balagovi}.
}.
As topologically nontrivial configurations,
't Hooft-Polyakov (TP) monopole configurations
were introduced \cite{Balachandran:2003ay,AIN3},
and an index theorem for those backgrounds 
was formulated
by introducing a projection operator
\cite{AIM}.
This index theorem was further extended to general 
configurations,
which enabled us to define 
all of the topological sectors in a single theory
\cite{AIM,Aoki:2008qta}.

While our formulation has been given so far to fermionic fields with 
the fundamental representation of the gauge group,
the matrix models of superstrings,
such as the IIB matrix model,
have fermions with the adjoint representation.
It is then desirable to provide formulations for the adjoint matter.
Since it is a highly delicate problem 
to formulate GW relations in each concrete case,
we will study it in this paper.
We further extend our formulation to configurations where
the $U(\sum_p k_p)$ gauge symmetry is  broken down to 
$\prod_p U(k_p)$,
which seem phenomenologically interesting.

The formulations using the GW relation provide a firm foundation for 
studying the above mentioned mechanism of obtaining chiral fermions 
by embedding topological configurations in the extra dimensions.
Indeed, the GW relation ensures the existence of chiral zero modes against
any perturbations  
since the index is a topological quantity.
However, one should study carefully whether the chiral zero modes
in the extra dimensions really give chiral spectrum in our spacetime.
By considering TP monopole-type configurations, 
where the gauge symmetry is broken down to a smaller one,
bifundamental fermions are obtained 
from an adjoint one, but fields with the conjugate representations
arise in pairs.
Whether they give chiral spectrum in our spacetime in total is
a problem and will be also
discussed in this paper.

In section \ref{sec:GWadj}, we formulate the GW relation 
for matter in the adjoint representation of the gauge group.
In section \ref{sec:monopole},
we introduce TP monopole configurations and provide 
the index theorem for those backgrounds.
We then extend it to general configurations 
in section \ref{sec:genecon211}.
We study configurations
with $U(\sum_p k_p)/\prod_p U(k_p)$ in section \ref{sec:config_k}.
In section \ref{sec:embedding}, we discuss whether topological configurations
in the extra dimensions
really provide chiral fermions in our spacetime.
Section \ref{sec:conclusion} is devoted to conclusions and discussions.
In appendix \ref{sec:comlim_D_TC},
we show calculations for taking
the commutative limits of the Dirac operator and the topological charge.
In appendix \ref{sec:genecon_Uk},
we study general configurations with $U(\sum_p k_p)/\prod_p U(k_p)$.
In appendix \ref{sec:chargecon},
we study the charge conjugation and the Majorana condition in 
ten dimensions in detail.

\section{GW relation on fuzzy $S^2$ with adjoint matter}
\label{sec:GWadj}
\setcounter{equation}{0}

In this section, we provide a Ginsparg-Wilson (GW) Dirac operator
and an index theorem for matter in the adjoint representation of
the gauge group,
by following the general prescription given in \cite{AIN2}.

Noncommutative coordinates of 
a fuzzy 2-sphere are given by
$x_i =\alpha L_i$, 
where $\alpha$ is a noncommutative parameter,
and $L_i$ is the $n$-dimensional irreducible
representation matrix of the $SU(2)$ algebra.
One then has the relation
$
(x_i)^2
=\alpha^2 \frac{n^2-1}{4}\id_n 
= \rho^2 \id_n 
$,
where 
$\rho=\alpha \sqrt{(n^2-1)/4}$ 
expresses the radius of the sphere.
The commutative limit is taken by 
$\alpha \to0, n \to \infty$
with $\rho$ fixed.

In our formulation of the GW relation,
we first define two chirality operators as\footnote{In the 
case of fundamental matter, we took
$\Gamma = a(\sigma_i L_i^R -\frac{1}{2})$ instead of (\ref{def_G}),
where $a=2/n$ is
a noncommutative analog of the lattice-spacing.
$\hat\Gamma$ was identical with (\ref{def_hatG}).}
\beqa
\Gamma =  \frac{H_r}{\sqrt{(H_r)^2}}  &,&  \ \ 
H_r=\sigma_i A^R_i -\frac{1}{2}  \ ,
\label{def_G}  \\
\hat\Gamma = \frac{H_l}{\sqrt{(H_l)^2}}   &,& \ \ 
H_l=\sigma_i A^L_i +\frac{1}{2}  \ ,
\label{def_hatG} 
\eeqa
with covariant coordinates
\beq
A_{i} = L_{i}+\rho a_{i} \ .
\label{defcovder}
\eeq
The superscript $R$ ($L$) in $A_i^R$ ($A_i^L$) 
means that this operator acts from the right (left) on matrices:
$A^L M \equiv AM , \ A^R M \equiv MA$.
The matrices $\sigma_i$ are the Pauli matrices acting on the spinor indices,
and the matrices $a_i$ in (\ref{defcovder}) represent 
the gauge fields.
$U(k)$ gauge symmetry is introduced
 by taking
$L_i = L_i \otimes \id_k$
and
$a_i = a_i^a t^a$ in (\ref{defcovder}),
where $t^a$'s are the generators of $U(k)$
and $a_i^a$'s are functions of the coordinates $L_i$.

The gauge transformation for the fermionic fields $\psi$ 
in the adjoint representation is given by
\beq
\psi \rightarrow U \psi U^\dagger \ ,
\eeq
where $U$ is $U(nk)$ matrices.
The gauge field $a_i$ is transformed as
$
a_i \rightarrow U a_i U^\dag +\frac{1}{\rho} (U L_i U^\dag-L_i) \ ,
$
so that the covariant coordinate $A_i$ is transformed as
\beq
A_i \rightarrow U A_i U^\dagger \ .
\eeq
Hence, both $\Gamma \psi$ and $\hat{\Gamma} \psi$ 
are transformed covariantly 
as $\Gamma \psi \rightarrow U \Gamma \psi U^\dagger$
and $\hat\Gamma \psi \rightarrow U \hat\Gamma \psi U^\dagger$,
where a relation
$(AB)^R \psi =  B^R A^R \psi = \psi AB$
was used.

The chirality operators (\ref{def_G}) and (\ref{def_hatG}) satisfy
\begin{equation}
\Gamma^\dagger=\Gamma \ , \ \ \
\hat\Gamma^\dagger=\hat\Gamma \ , \ \ \
\Gamma^2=\hat\Gamma^2=1 \ .
\end{equation}
In the commutative limit, both $\Gamma$ and $\hat\Gamma$
become the chirality operator on the commutative 2-sphere,
$\gamma = n_i \sigma_i$,
where $n_i = x_i/\rho$ is a unit vector.

We then define a GW Dirac operator as
\begin{equation}
D_{\rm GW} = -a^{-1}\Gamma (1- \Gamma \hat{\Gamma}) \ ,
\label{defDGW}
\end{equation}
where $a=2/n$ is a noncommutative analog of the lattice spacing.
By the definition,
a GW relation  
\begin{equation}
\Gamma D_{\rm GW}+D_{\rm GW} \hat{\Gamma}=0 
\label{GWrelation}
\end{equation}
is satisfied. Hence, the index, {\it i.e.}, the difference of the numbers 
of the chiral zero modes, is given by
the trace of the chirality operators as
\begin{equation}
{\rm{index}}(D_{\rm GW})=\frac{1}{2}
\CTr[\Gamma +\hat{\Gamma}] \ , 
\label{ITtrivial}
\end{equation}
where $\CTr$ is the trace over the whole configuration space,
that is, over the spinor index, the gauge group space,
and the matrix space representing the coordinates.
Since the definition of $\Gamma$ and $\hat\Gamma$ depends on the gauge
fields $a_i$, 
the right-hand side (rhs) of (\ref{ITtrivial}) is a functional of the
gauge field configurations.
It also takes only integer values.
It then gives a noncommutative
generalization of the topological charge of the gauge field backgrounds.
Thus, eq.(\ref{ITtrivial}) gives
an index theorem on the fuzzy 2-sphere.

In the commutative limit, the GW Dirac operator (\ref{defDGW}) becomes
\begin{equation}
D_{\rm GW} \rightarrow 
\sigma_i ({\cal L}_i + \rho P_{ij} \tilde{a}_j ) +1 \ ,
\label{DGWcom}
\end{equation}
as will be shown in appendix \ref{sec:comlim_D_TC}.
Here  ${\cal L}_i=-i\epsilon_{ijk} x_j \partial_k$
is the derivative operator along the Killing vectors on the sphere,
$\tilde{a}_i$ is the adjoint operator of $a_i$, {\it i.e.},
$\tilde{a}_i \psi = [a_i , \psi]$,
and $P_{ij}=\delta_{ij}-n_i n_j$ 
is the projector to the tangential directions on the sphere. 
The gauge fields $a_i$ can be 
decomposed into the tangential components on the 
sphere $a_i'$ and the normal component $\phi$ as
\begin{eqnarray}
&&\left\{
\begin{array}{lll}
a_i'&=& \epsilon_{ijk}n_j a_k \ , \\
\phi&=&n_i a_i \ ,
\end{array}
\right. \label{decomposeto}\\
&\Leftrightarrow& a_i = -\epsilon_{ijk}n_j a_k' + n_i \phi \ .
\label{decomposefrom}
\end{eqnarray}
The normal component $\phi$ is a scalar field on the sphere.
The operator (\ref{DGWcom}) is 
the Dirac operator of the adjoint matter on the commutative 2-sphere
without a coupling to the scalar field $\phi$.
The absence of the Yukawa coupling is reasonable
since such a coupling would violate the chiral symmetry
on the sphere and contradict with the GW relation.

The commutative limit of the topological charge,
the rhs of (\ref{ITtrivial}), becomes
\beq
\frac{1}{2} \CTr [\Gamma + \hat\Gamma]
\to
-\rho^2 
 \int \frac{d\Omega}{4\pi} 
\tr (\epsilon_{ijk} n_{k}F_{ij})
+\rho^2 
 \int \frac{d\Omega}{4\pi} 
\tr (\epsilon_{ijk} n_{k}F_{ij})  \ ,
\label{com_top_char_trivial}
\eeq
as shown in appendix \ref{sec:comlim_D_TC}.
Here $\tr$ is the trace over the gauge group space,
and
the field strength $F_{ij}$ is defined as
$F_{ij}= \partial_i a_j'-\partial_j a_i'-i[a_i',a_j']$
with $a'_i$ 
given in (\ref{decomposeto}).
The first and the second terms on the rhs of (\ref{com_top_char_trivial})
come from $\CTr [\Gamma]$ and $\CTr [\hat\Gamma]$, respectively.
Each term gives the integral of 
the 1st Chern character on the commutative 2-sphere. 
They cancel each other and vanish for any gauge field configurations,
which is appropriate since we now consider the adjoint matter.

In summary, our formulation manifestly has
the gauge invariance and the $SO(3)$ Poincare 
invariance on the fuzzy 2-sphere.
Because of the GW relation, the index theorem (\ref{ITtrivial}) is satisfied, 
and the topological charge, the rhs of (\ref{ITtrivial}), takes only integer values.
The commutative limits of the chirality operators, the Dirac operator, and 
the topological charge have the correct forms.

\section{TP Monopole configurations}
\label{sec:monopole}
\setcounter{equation}{0}

As topologically nontrivial configurations
in the $U(2)$ gauge theory on the fuzzy 2-sphere,
the following configurations were provided 
\cite{Balachandran:2003ay,AIN3}:
\begin{equation}
A_i=
\begin{pmatrix}
 L_i^{(n+m)} & \cr
& L_i^{(n-m)} \cr
\end{pmatrix} 
\label{LnpmLnmm} \ , 
\end{equation}
where $A_i$ is the covariant coordinate (\ref{defcovder}),
and $L_i^{(n\pm m)}$ are the $(n\pm m)$-dimensional 
irreducible representations of the $SU(2)$ algebra.
The $m=0$ case corresponds to two coincident fuzzy 2-spheres,
whose effective action is the $U(2)$ gauge theory.
The cases with  general $m$ correspond to two fuzzy 2-spheres 
with different radii.
They correspond
to the 't Hooft-Polyakov (TP) monopole configurations with magnetic charge $-|m|$,
where the $U(2)$ gauge symmetry  is broken down
to $U(1) \times U(1)$.

For the $m=1$ case,
(\ref{LnpmLnmm}) is unitarily equivalent to 
\begin{equation}
A_i \doteq L_i^{(n)} \otimes \id_2 +
\id_{n} \otimes \frac{\tau_i}{2} \ .
\label{AiLtau}
\end{equation}
Comparing with (\ref{defcovder}), the gauge field is 
\begin{equation}
a_i=\frac{1}{\rho}\id_{n} \otimes \frac{\tau_i}{2} \ .
\end{equation}
By taking the commutative limit 
and making the decomposition (\ref{decomposeto}), 
we obtain
\beqa
a_i^{\prime a} &=& \frac{1}{\rho} \epsilon_{ija} n_j \ , \n
\phi^a &=& \frac{1}{\rho} n_a \ ,
\eeqa
which is precisely the TP monopole configuration \cite{AIN3}.

We now define projection operators
$P^{(\pm)}$ to pick up
the $(n \pm |m|)$-dimensional spaces 
that the operator (\ref{LnpmLnmm}) acts.
It is written as
\beq
P^{(\pm)}
= \frac{1}{2}(1\pm T) \ ,
\eeq
with
\begin{eqnarray}
T &=& \frac{2}{n|m|}\left(A_i^2 - \frac{n^2+m^2-1}{4}\right) 
\label{defT}\\
  &=& \frac{m}{|m|}\begin{pmatrix}
 \id_{n+m} & \cr
& -\id_{n-m} \cr
\end{pmatrix} \ .
\label{1npm1nmp}
\end{eqnarray}

Since $T$ commutes with the chirality operators and the Dirac operator,
the index theorem (\ref{ITtrivial}) is satisfied in each space projected
by $P^{(\pm)}$ as
\begin{equation}
{\rm{index}}(P^{(\pm)L}P^{(\pm)R}D_{\rm GW})
=\frac{1}{2}\CTr[P^{(\pm)L}P^{(\pm)R}(\Gamma +\hat{\Gamma})] \ , 
\label{ITmonopole}
\end{equation}
where the superscript $L$ ($R$) means that 
the operator acts from the left (right)
on matrices as before.
The $\pm$ signs in $P^{(\pm)L}$ and $P^{(\pm)R}$ do not necessarily coincide.
Each sign combination picks up one of the following blocks in the fermionic field $\psi$
in the adjoint representation:
\beq
\psi=
\begin{pmatrix}
\psi^{(++)} & \psi^{(+-)} \cr
\psi^{(-+)} & \psi^{(--)} \cr
\end{pmatrix} 
\label{psi_block_decompose}
\eeq
for $m >0$,
if we decompose $\psi$ into the blocks in the same way as (\ref{LnpmLnmm}).
The signs in (\ref{psi_block_decompose}) should be reversed for $m < 0$.

For the backgrounds (\ref{LnpmLnmm}),
the rhs of (\ref{ITmonopole}) becomes
\beq
\frac{1}{2}\CTr[P^{(\pm)L}P^{(\pm)R}(\Gamma +\hat{\Gamma})]
=\left\{
\begin{array}{ll}
0 & {\rm for} \ \ \psi^{(++)}, \psi^{(--)}\\
-2|m| & {\rm for} \ \ \psi^{(+-)} \\
2|m| & {\rm for} \ \ \psi^{(-+)} \\ 
\end{array}
\right.
\label{PLPRGplushatGmono}
\eeq
as shown by the following calculations:
For (\ref{LnpmLnmm}), the chirality operator $\hat\Gamma$ becomes
\beq
\hat\Gamma=
\begin{pmatrix}
\frac{2}{n+m}(\sigma \cdot L^{(n+m)}+\frac{1}{2})& \cr
& \frac{2}{n-m}(\sigma \cdot L^{(n-m)}+\frac{1}{2})  \cr
\end{pmatrix} \ .
\eeq
Since the terms with $\sigma \cdot L$ vanish after taking the trace,
we obtain
\beqa
\CTr[P^{(\pm)L}P^{(\pm)R}\hat\Gamma]
&=&\Tr_{L,\sigma}[P^{(\pm)L}\hat\Gamma] \cdot \Tr_{R}[P^{(\pm)R}] \n
&=&\frac{1}{n \pm |m|} 2(n \pm |m|) \cdot (n \pm |m|) \n
&=&2(n \pm |m|) \ ,
\label{PLPRhatG}
\eeqa
where $\Tr_{L,\sigma}$ is the trace over the space on which
$A_i^L$ and $\sigma_i$ act,
and $\Tr_{R}$ is the trace over the space on which $A_i^R$  act.
The $\pm$ sign in the last line refers to that in $P^{(\pm)R}$.
Similarly, we can show
\beq
\CTr[P^{(\pm)L}P^{(\pm)R}\Gamma]
=-2(n \pm |m|) \ ,
\label{PLPRG}
\eeq
where the $\pm$ sign in the rhs refers to that in $P^{(\pm)L}$.
By adding (\ref{PLPRhatG}) and (\ref{PLPRG}), 
we obtain (\ref{PLPRGplushatGmono}).

We now give an interpretation for (\ref{PLPRGplushatGmono}).
In the representation (\ref{defcovder}),
 (\ref{defT}) is written as
\begin{equation}
T  =\frac{2}{n|m|}\left(\rho \{ L_i , a_i \}+ \rho^2 a_i^2-\frac{m^2}{4}\right) \ .
\label{Tother}
\end{equation}
In the commutative limit, $T$ becomes
$\frac{2\rho}{|m|} \phi$ where $\phi$ is the scalar field 
defined in (\ref{decomposeto}).
It is also normalized as $T^2=\id_{2n}$. 
Then, $T$ corresponds to a normalized scalar field.
Recalling that the TP monopole configuration breaks
the $SU(2)$ gauge symmetry down to $U(1)$,
$T$ is the generator of this unbroken $U(1)$, 
the electric charge operator of the unbroken $U(1)$.
(The $U(1)$ of $U(2) \simeq  SU(2) \times U(1)$ is ignored
since it is decoupled in the commutative limit.)
By the gauge symmetry braking $SU(2)/U(1)$,
fields with various electric charges of the unbroken $U(1)$ arise.
Equation (\ref{ITmonopole}) gives the index theorem for each field.

For instance, $\psi^{(++)}$ in (\ref{psi_block_decompose}) is in the
adjoint representation of the unbroken $U(1)$ with electric charge $+1/2$,
and it has a vanishing index.
On the other hand, $\psi^{(+-)}$ is in the bifundamental representation
of the unbroken $U(1)$ with charge $+1/2$ and $-1/2$, 
that is, the fundamental representation with charge $+1$.
It therefore has the index $-2|m|$.
Although the whole fermionic field $\psi$ has a vanishing index 
since it is in the adjoint representation,
the field in each projected block can have nonzero index.
As was shown in (\ref{com_top_char_trivial}),
topological charge is an analog of the 1st Chern character,
which is proportional to the electric charge of the matter.
Then,
 $\psi^{(+-)}$ and $\psi^{(-+)}$, having the opposite electric charge,
have the opposite topological charge and the opposite index.

We finally give two comments.
First,
we can define a topological charge
multiplied by the electric charge, such as
\beq
\frac{1}{16}{\cal T}r[(T^L-T^R)(\Gamma +\hat{\Gamma})] \ ,
\label{topchar_TL-TR_G+hG}
\eeq
so that 
contributions from the blocks in (\ref{psi_block_decompose}) do not cancel but are added.
By using the result (\ref{PLPRGplushatGmono}),
(\ref{topchar_TL-TR_G+hG}) becomes $-|m|$ for the backgrounds (\ref{LnpmLnmm}),
which agrees with the topological charge of the TP monopoles.
We will develop this argument further in the next section.

Second, 
as seen above,
fermions in the conjugate representations under the unbroken gauge group have opposite indices
if one considers topological configurations in two dimensions,
or more generally, in $2$ (mod $4$) dimensions.
We can then expect that
by embedding these configurations in the extra dimensions,
chiral spectrum is obtained in our spacetime
in low energy effective theory.
We will discuss this issue in section \ref{sec:embedding}.

\section{General configurations with $U(2)/U(1)^2$}
\label{sec:genecon211}
\setcounter{equation}{0}

We now extend the formulation 
in the previous section to general configurations
where the $U(2)$ gauge group is broken down to $U(1) \times U(1)$
through the Higgs mechanism, {\it i.e.},  a nonzero 
vacuum expectation value of the scalar field.
This will enable us to survey the whole configuration space
with all topological sectors.

Since the definition of the electric charge operator $T$ in (\ref{defT})
was specific to the backgrounds (\ref{LnpmLnmm}),
we first generalize it as
\beq
T' = \frac{(A_i)^2-\frac{n^2-1}{4}}
{\sqrt{\left[ (A_i)^2-\frac{n^2-1}{4} \right]^2}} \ .
\label{defTgen}
\eeq
This is valid for general configurations $A_i$
unless the denominator has zero modes.
For the configurations (\ref{LnpmLnmm}),
$T'$ reduces to the previous one (\ref{1npm1nmp}).
For general configurations
\beq
(T')^\dag =T' \ , \ (T')^2 = 1 \ 
\eeq
are satisfied.
The commutative limit of $T'$ becomes the normalized scalar field as
\beq
T' \to 2 \phi'=2\phi'^a \frac{\tau^a}{2} \ ,
\label{comlimTprime}
\eeq
where $\phi'$ is normalized as $\sum_a (\phi'^a)^2 = 1$.

We next define modified chirality operators as
\beqa
\Gamma'_r &=& \frac{\{T'^R, \Gamma \}}{\sqrt{\{T'^R, \Gamma \}^2}} \ ,
\label{def_Gr} \\
\hat\Gamma'_r &=& T'^R \hat\Gamma \ , 
\label{def_hatGr} \\
\Gamma'_l &=& T'^L \Gamma \ , 
\label{def_Gl} \\
\hat\Gamma'_l &=& \frac{\{T'^L, \hat\Gamma \}}{\sqrt{\{T'^L, \hat\Gamma \}^2}} \ , 
\label{def_hatGl} 
\eeqa
where $\Gamma$ and $\hat\Gamma$ are defined in
(\ref{def_G}) and (\ref{def_hatG}).
The superscript $R$ ($L$) in $T'^R$ ($T'^L)$ means that this operator acts 
from right (left) on matrices.
The chirality operators satisfy the relations
\beqa
&&(\Gamma'_r)^\dag = \Gamma'_r \ , \ \ \
(\hat\Gamma'_r)^\dag =\hat\Gamma'_r \ , \ \ \
(\Gamma'_r)^2 = (\hat\Gamma'_r)^2 =1 \ ,\\
&&(\Gamma'_l)^\dag = \Gamma'_l \ , \ \ \
(\hat\Gamma'_l)^\dag =\hat\Gamma'_l \ , \ \ \
(\Gamma'_l)^2 = (\hat\Gamma'_l)^2 =1 \ .
\eeqa
Since the chirality operators are weighted by the electric charge 
operator $T'$, 
the commutative limits of $\Gamma'_r$ and $\hat\Gamma'_r$
become $\gamma'_r = t^R \gamma$,
and those of $\Gamma'_l$ and $\hat\Gamma'_l$
become $\gamma'_l = t^L \gamma$.
Here $t$ is the electric charge operator of
the unbroken $U(1)$ gauge group,
the superscript $R$ ($L$) means that the operator acts from right (left)
in the gauge group space, and $\gamma=n\cdot\sigma$ is the
chirality operator on the 2-sphere.

We then define modified GW Dirac operators as
\beqa
D'_r &=& -a^{-1} \Gamma'_r (1 - \Gamma'_r \hat\Gamma'_r) \ ,
\label{def_DGWgenr} \\
D'_l &=& -a^{-1} \Gamma'_l (1 - \Gamma'_l \hat\Gamma'_l) \ .
\label{def_DGWgenl}
\eeqa
By definition, 
these Dirac operators satisfy GW relations
\beqa
\Gamma'_r D'_r + D'_r \hat\Gamma'_r &=& 0 \ , 
\label{GWrel_Dpr}\\
\Gamma'_l D'_l + D'_l \hat\Gamma'_l &=& 0 \ .
\label{GWrel_Dpl}
\eeqa
Then, index theorems
\beqa
{\rm index}(D'_r) &=& 
\frac{1}{2} {\cal T}r [\Gamma'_r + \hat\Gamma'_r] \ ,
\label{ITgenr} \\
{\rm index}(D'_l) &=& 
\frac{1}{2} {\cal T}r [\Gamma'_l + \hat\Gamma'_l] \ ,
\label{ITgenl}
\eeqa
are satisfied as well. 
By using the rhs of (\ref{ITgenr}) and (\ref{ITgenl}),
we can also define a topological charge 
\beq
\frac{1}{16} \CTr[\Gamma'_l + \hat\Gamma'_l-\Gamma'_r - \hat\Gamma'_r] \ ,
\label{TC_gen_l-r}
\eeq
which is a generalization of (\ref{topchar_TL-TR_G+hG}).

For the configurations (\ref{LnpmLnmm}),
since the generalized electric charge operator (\ref{defTgen})
reduces to the previous one (\ref{1npm1nmp}),
we can calculate the rhs of (\ref{ITgenr}) and  (\ref{ITgenl})
as we did below (\ref{PLPRGplushatGmono}), giving
\beqa
\frac{1}{2} {\cal T}r [\Gamma'_r + \hat\Gamma'_r] &=& 4|m| \ ,
\label{topchar_genr_mono} \\
\frac{1}{2} {\cal T}r [\Gamma'_l + \hat\Gamma'_l] &=& -4|m| \ .
\label{topchar_genl_mono}
\eeqa
In (\ref{PLPRGplushatGmono}), $\psi^{(+-)}$ and $\psi^{(-+)}$ have 
index $-2|m|$ and $2|m|$, respectively.
However, since the chirality operators $\Gamma'_r$  and  $\hat\Gamma'_r$ are multiplied 
by $-1$ for $\psi^{(+-)}$, we obtain (\ref{topchar_genr_mono}).
Equation (\ref{topchar_genl_mono}) is obtained similarly.
From (\ref{topchar_genr_mono}) and (\ref{topchar_genl_mono}),
the topological charge (\ref{TC_gen_l-r}) becomes $-|m|$,
as expected since (\ref{topchar_TL-TR_G+hG}) gave $-|m|$.

In the commutative limit, the GW Dirac operator (\ref{def_DGWgenr}) becomes
\beq
D'_r \ \to \frac{1}{2} \{2\phi'^R , (\sigma_i {\cal L}_i +1) \}
+ \frac{1}{2} \{2\phi'^R  , \rho \sigma_i P_{ij} a_j^L \} 
-\frac{1}{2} \{2\phi'^R  , \rho \sigma_i P_{ij} a_j^R \}  \ ,
\label{com_lim_Dpr}
\eeq
where the superscript $R$ ($L$) means that the operator acts 
from right (left) in the gauge group space:
$\phi'^R = \phi'^a(\Omega)\frac{(\tau^a)^R}{2}$, {\it etc}.
In the $\phi'^a(\Omega) = (0,0,1)$ gauge, (\ref{com_lim_Dpr}) becomes
\beq
(\tau^3)^R \biggl(\sigma_i {\cal L}_i +1
+ \rho \sigma_i P_{ij} 
\Bigl(a_j^3 \frac{\tilde\tau^3}{2} + a_j^1 \frac{(\tau^1)^L}{2}
+ a_j^2 \frac{(\tau^2)^L}{2}\Bigr)\biggr) 
\equiv D'_{r,{\rm com}}\ ,
\eeq
where $\tilde\tau^3$ means the adjoint operator of $\tau^3$.
This Dirac operator indeed has the adjoint coupling of the unbroken $U(1)$
gauge field $a_i^3$.
It also satisfies a chiral relation 
\beq
\{D'_{r,{\rm com}}  ,  \gamma'_r \} = 0 \ ,
\eeq
with $ \gamma'_r = (\tau^3)^R \gamma$ the chirality operator
multiplied by the unbroken $U(1)$ charge,
as expected from the GW relation (\ref{GWrel_Dpr}).
The same arguments hold also for $D'_l$.

Our remarkable result is that,
by the same calculations in (\ref{com_top_char_trivial}),
the commutative limit of the rhs in (\ref{ITgenr}) becomes
\beq
\frac{1}{2} {\cal T}r [\Gamma'_r + \hat\Gamma'_r]
\to
-4\frac{\rho^2}{8\pi}\int d\Omega \epsilon_{ijk}
n_i \Bigl( \phi'^a F_{jk}^a 
- \epsilon_{abc} \phi'^a (D_j \phi')^b (D_k \phi')^c \Bigr) \ ,
\label{comlim_TCgenr} 
\eeq
where $F_{jk}=F_{jk}^a \tau^a/2$ 
is the field strength defined as
$F_{jk}= \partial_j a_k'-\partial_k a_j'-i[a_j',a_k']$,
and $D_{j}$ 
is the covariant derivative defined as
$D_{j} =\partial_{j}  -i[a'_{j} , ~~]$, with
$a'_j$ given in (\ref{decomposeto}).
As $\CTr(\hat\Gamma)$ gave the second term in the rhs of (\ref{com_top_char_trivial}),
$\CTr(\hat\Gamma'_r)$ gives a similar term,
but with $\Tr_R(\id)=2n$ replaced by  $\Tr_R(T'^R) \sim 2m$,
giving an extra $1/n$ factor.
Then, $\CTr(\hat\Gamma'_r)$  does not contribute to the commutative limit.
On the other hand, $\CTr(\Gamma'_r)$ gives a similar term as the first term in
the rhs of (\ref{com_top_char_trivial}),
but with the $T'^R$ in the same trace.
Moreover, as shown in ref.\cite{Aoki:2008qta},
the denominator in (\ref{def_Gr}) yields
the second term on the rhs of (\ref{comlim_TCgenr}).

Similarly, we obtain
\beq
\frac{1}{2} {\cal T}r [\Gamma'_l + \hat\Gamma'_l]
\to
4\frac{\rho^2}{8\pi}\int d\Omega \epsilon_{ijk}
n_i \Bigl( \phi'^a F_{jk}^a 
- \epsilon_{abc} \phi'^a (D_j \phi')^b (D_k \phi')^c \Bigr) \ .
\label{comlim_TCgenl}
\eeq
Equations (\ref{comlim_TCgenr}) and (\ref{comlim_TCgenl})
are precisely the topological charge given by 't~Hooft
\cite{'tHooft:1974qc},
multiplied by $\mp 4$, respectively.
Since each of (\ref{comlim_TCgenr}) and (\ref{comlim_TCgenl}) has contributions
from $\psi^{(+-)}$ and $\psi^{(-+)}$,
and their electric charge is twice the usual case,
the result is multiplied by $\mp 4$.

\section{Configurations with $U(\sum_p k_p)/\prod_p U(k_p)$}
\label{sec:config_k}
\setcounter{equation}{0}

We now consider configurations as follows:
\begin{equation}
A_i=
\begin{pmatrix}
 L_i^{(n_1)}\otimes \id_{k_1} & & & \cr
& L_i^{(n_2)}\otimes \id_{k_2} & & \cr
& & \ddots & \cr
& & & L_i^{(n_h)}\otimes \id_{k_h} \cr
\end{pmatrix} 
\label{config_ka} \ , 
\end{equation}
where the gauge symmetry $U(\sum_{p=1}^h k_p)$, which the configurations
$A_i = L_i \otimes \id_{\sum_{p=1}^h k_p}$ would have, is broken down to
$\prod_{p=1}^h U(k_p)$.
They are a generalization of the configurations ({\ref{LnpmLnmm}}) with $U(2)/U(1)^2$.
They are phenomenologically attractive since they have gauge group
close to that of the standard model\footnote{
A phenomenological study based on such configurations was given in \cite{Grosse:2010zq}.}.
Such configurations are also used for embedding fiber bundles 
in matrix models \cite{ISTT}.
We here study whether index theorems can be formulated in these backgrounds as before.

We then define projection operators as
\beq
P_p =
\begin{pmatrix}
0_{\sum_{q=1}^{p-1} n_q k_q} & & \cr
& \id_{n_p k_p} & \cr
&& 0_{\sum_{q=p+1}^{h} n_q k_q}  \cr
\end{pmatrix} \ 
\label{projection_confka}
\eeq
for $p=1,\ldots , h$,
which pick up the $p$th block with dimensions $n_p k_p$.
Since the projection operators (\ref{projection_confka}) commute with
the chirality operators and the Dirac operator,
the index theorem (\ref{ITtrivial}) is satisfied in each projected space
as 
\begin{equation}
{\rm{index}}(P_p^L P_q^R D_{\rm GW})=
\frac{1}{2} \CTr[P_p^L P_q^R(\Gamma +\hat{\Gamma})] \ 
\label{ITconfka}
\end{equation}
for $1 \le p,q \le h$.
Here $\Gamma$, $\hat{\Gamma}$ and $D_{\rm GW}$ are defined in
(\ref{def_G}), (\ref{def_hatG}) and (\ref{defDGW}), and
the superscript $L$ ($R$) means that the operator acts from the left (right).

For the backgrounds (\ref{config_ka}),
the rhs of (\ref{ITconfka}) becomes
\beq
\frac{1}{2} \CTr[P_p^L P_q^R(\Gamma +\hat{\Gamma})] 
= - k_p k_q (n_p-n_q) \ ,
\eeq
by following the same calculations below (\ref{PLPRGplushatGmono}).
For $h=2$ and $k_1=k_2=1$, this reproduces the previous result 
(\ref{PLPRGplushatGmono}).
Since the field projected by $P_p^L$ and $P_q^R$ is in the bifundamental representation $(k_p,\bar{k_q})$
of the unbroken gauge group $U(k_p)\times U(k_q)$,
its index is multiplied by $k_p k_q$.

We can also extend the formulation to general configurations.
As in (\ref{defTgen}), we define electric charge operators 
of the unbroken $U(1)$'s as
\beq
T'_p = \frac{(A_i)^2-c_p}
{\sqrt{\left[ (A_i)^2-c_p \right]^2}} \ 
\label{defTgenka}
\eeq
for $p = 1, \ldots, h-1$.
The numbers $c_p$ are taken between
\[
\frac{n_p^2-1}{4} > c_p > \frac{n_{p+1}^2-1}{4} \ ,
\]
where we assume $n_1 > n_2 > \cdots  > n_h$.
For the configurations (\ref{config_ka}), $T'_p$ becomes
\beq
\begin{pmatrix}
\id_{\sum_{q=1}^p n_q k_q} & \cr
& -\id_{\sum_{q=p+1}^h n_q k_q} \cr
\end{pmatrix} \ .
\eeq
They are the generators of $U(1)$'s contained in
the unbroken gauge group $\prod_p U(k_p)$.
Note that there exist the grand unified theory monopoles
when a simple gauge group is broken down to
a smaller group containing $U(1)$ factors.
We then define modified chirality operators as 
(\ref{def_Gr})--(\ref{def_hatGl})
for each $T'_p$ with $p = 1, \ldots, h-1$.
GW Dirac operators, GW relations, and index theorems are defined
as (\ref{def_DGWgenr})--(\ref{ITgenl}).
As we show in appendix \ref{sec:genecon_Uk},
the commutative limits of the GW Dirac operators and the topological charges
have similar forms as 
(\ref{com_lim_Dpr})--(\ref{comlim_TCgenl}).

\section{Embeddings in IIB matrix model}
\label{sec:embedding}
\setcounter{equation}{0}

As we mentioned in the Introduction, 
when topologically nontrivial configurations are embedded in the extra dimensions
in the matrix model formulations of superstring theory, such as the IIB matrix model,
chiral fermions can be obtained in our spacetime.
In this section, we discuss whether this mechanism really works or not. 

\subsection{$M^4 \times X^n \subset M^{4+n}$}

Let us first consider general cases,
theories in $(4+n)$-dimensional Minkowski space $M^{4+n}$, 
compactified to $n$-dimensional space $X^n$ with Euclidean signature,
while $M^4$ is our spacetime with Lorentzian signature.
We then embed $n$-dimensional topological configurations in $X^n$.
In particular, we assume configurations of the TP monopole type, 
where the gauge symmetry is broken down,
which yields fields that are in the conjugate representations under the 
unbroken gauge group.
We now denote them as $\psi^{(r)}$ and $\psi^{(\bar{r})}$,
which correspond to $\psi^{(+-)}$ and $\psi^{(-+)}$ in (\ref{psi_block_decompose}).

For $n=2$ (mod $4$),
as we mentioned at the end of section \ref{sec:monopole},
topological charge becomes an analog of the $l$th Chern character with $l=n/2$ an odd integer,
which gives $\psi^{(r)}$ and $\psi^{(\bar{r})}$ opposite indices. 
We denote the corresponding chiral zero modes as $\psi^{(r)}_R$ and $\psi^{(\bar{r})}_L$,
where the subscripts $R$ and $L$ stand for the chirality.
(Choosing $\psi^{(r)}_L$ and $\psi^{(\bar{r})}_R$ instead would give the identical
results below.)
Taking spinors $\varphi$ in $M^4$ as well,
we obtain four possible Weyl spinors as follows:
\beqa
\varphi_R \otimes \psi^{(r)}_R \ , \label{d2RR}\\
\varphi_L \otimes \psi^{(\bar{r})}_L \ , \label{d2LL}\\
\varphi_L \otimes \psi^{(r)}_R \ , \label{d2LR}\\
\varphi_R \otimes \psi^{(\bar{r})}_L \ .\label{d2RL}
\eeqa
The spinors (\ref{d2RR}) and (\ref{d2LL}) are in the charge conjugate representations 
to each other.
So are (\ref{d2LR}) and (\ref{d2RL}).
Here one should note that Weyl spinors in Lorentzian and Euclidean spaces
are as shown in table \ref{table:Weylrep}.
\begin{table}[t]
\begin{center}
\begin{tabular}{l|l|l}
& $SO(d-1,1)$ & $SO(d)$ \\ \hline
d=0 (mod $4$) & Complex & Self-conjugate\\ 
d=2 (mod $4$) & Self-conjugate & Complex 
\end{tabular}
\end{center}
\caption{Weyl representations of $SO(d-1,1)$ and $SO(d)$.}
\label{table:Weylrep}
\end{table}

If we consider chiral theories in $M^{4+n}$ originally,
(\ref{d2RR}) and (\ref{d2LL}) are chosen.
(Choosing (\ref{d2LR}) and (\ref{d2RL}) would give the identical results.)
Since $\varphi_R$ in (\ref{d2RR}) and $\varphi_L$ in (\ref{d2LL})
are in the different representations of the gauge group,
we obtain chiral spectrum in $M^4$,
although we have a doubling of (\ref{d2RR}) and (\ref{d2LL}).
If we further impose the Majorana condition in $M^{4+n}$,
which is possible for $4+n=2$ (mod $8$),
(\ref{d2RR}) and (\ref{d2LL}) are identified
and the doubling problem is resolved.

On the contrary, for $n=0$ (mod $4$),
topological configurations give  $\psi^{(r)}$ 
and $\psi^{(\bar{r})}$ the same index. 
We denote the corresponding chiral zero modes 
as $\psi^{(r)}_R$ and $\psi^{(\bar{r})}_R$.
Taking spinors $\varphi$ in $M^4$ as well,
we obtain
\beqa
\varphi_R \otimes \psi^{(r)}_R \ , \label{d0Rr}\\
\varphi_L \otimes \psi^{(\bar{r})}_R \ , \label{d0Lbr}\\
\varphi_L \otimes \psi^{(r)}_R \ , \label{d0Lr}\\
\varphi_R \otimes \psi^{(\bar{r})}_R \ .\label{d0Rbr}
\eeqa
The spinors (\ref{d0Rr}) and (\ref{d0Lbr}) are in the charge conjugate representations.
So are (\ref{d0Lr}) and (\ref{d0Rbr}).
If we consider chiral theories in $M^{4+n}$ originally,
(\ref{d0Rr}) and (\ref{d0Rbr}) are chosen.
Since $\varphi_R$ in (\ref{d0Rr}) and $\varphi_R$ in (\ref{d0Rbr}) are in 
the conjugate representations of the gauge group to each other,
we are left with nonchiral spectrum in $M^4$.
Even if we consider the Majorana fermions in $M^{4+n}$ instead,
we obtain a nonchiral spectrum in $M^4$.

\subsection{$M^4 \times S^2 \times S^2$ in IIB matrix model}

We now move to the IIB matrix model.
The action of the IIB matrix model is given by
\beq
S_{\rm IIBMM}  =  -{1\over g^2}{\rm Tr}\left({1\over 4}[A_{M},A_{N}][A^{M},A^{N}]
+{1\over 2}\bar{\psi}\Gamma ^{M}[A_{M},\psi ]\right) \ ,
\label{IIBMMaction}
\eeq
where 
$A_{M}$ is a ten-dimensional vector, $\psi$ is a ten-dimensional
Majorana-Weyl spinor\footnote{
They are Wick rotated to the $SO(10)$ vector and spinor.
In this paper, however, we use Lorentzian notation, such as $M^{10}$,
since we discuss spinors.}
and  
they are also traceless Hermitian matrices. 
Since the action is written in terms of the commutators,
matter in the adjoint representation appears naturally.

As an application of what we studied about the fuzzy 2-sphere in this paper,
let us consider a compactification to $M^4 \times S^2 \times S^2$ and
an embedding of the following configurations:
\beqa
A_\mu &=& x_\mu \otimes \id_{n^1_1 n^2_1  + n^1_2 n^2_2 } \ , \n
A_i &=& \id \otimes 
\begin{pmatrix}
L_i^{(n^1_1)} \otimes \id_{n^2_1}  & \cr
& L_i^{(n^1_2)} \otimes \id_{n^2_2}  \cr
\end{pmatrix} 
 \ , \n
A_j &=&  \id \otimes
\begin{pmatrix}
\id_{n^1_1}  \otimes L_j^{(n^2_1)}  & \cr
&  \id_{n^1_2}  \otimes L_j^{(n^2_2)}  \cr
\end{pmatrix} 
\ ,
\label{conf_embed_in_IIBMM}
\eeqa
where $\mu=0,1,2,3$, $i=4,5,6$ and $j=7,8,9$.
$x_\mu$ is our spacetime background.
Either commutative backgrounds as $[x_\mu, x_\nu]=0$
or noncommutative backgrounds as $[x_\mu, x_\nu]=i \theta_{\mu\nu}$
can be considered\footnote{
Fluctuations around the background (\ref{conf_embed_in_IIBMM}) provide matter fields. 
Expansions of the action (\ref{IIBMMaction}) give superficially renormalizable theories,
but with nonlocality such as noncommutativity.
The maximal supersymmetry possessed by the IIB matrix model might suppress 
peculiar properties caused by the nonlocality, such as the UV/IR mixing.
}.

The second factor in (\ref{conf_embed_in_IIBMM})\footnote{
Similar backgrounds were studied in \cite{Chatzistavrakidis:2009ix,Aoki:2009cv}.}
represents monopole configurations wrapping around $S^2 \times S^2$.
The off-diagonal blocks of matter,
$\psi^{(+-)}$ and $\psi^{(-+)}$ in (\ref{psi_block_decompose}),
are in the conjugate representations of the unbroken gauge group.
We now write them as $\psi^{(r)}$ and $\psi^{(\bar{r})}$.
Since the topological configurations in four-dimensional $S^2 \times S^2$
give $\psi^{(r)}$ and $\psi^{(\bar{r})}$ the same index,
we denote the corresponding chiral zero modes as
$\psi^{(r)}_R$ and $\psi^{(\bar{r})}_R$.

We now introduce the following Dirac gamma matrices in $M^{10}$, 
which are suitable for $M^4 \times S^2 \times S^2$:
\beqa
\Gamma_\mu &=& \gamma_\mu \otimes \id_2 \otimes \id_2 \otimes \sigma_3 \ , \n
\Gamma_i &=& \id_4 \otimes \sigma_i \otimes \id_2 \otimes \sigma_1 \ , \n
\Gamma_j &=& \id_4 \otimes \id_2 \otimes \sigma_j \otimes \sigma_2 \ ,
\label{gammaM4S2S2}
\eeqa
where $\gamma_\mu$ is the gamma matrices in $M^4$.
The second and the third factors act on spinors on $S^2 \times S^2$,
such as the chiral zero modes $\psi^{(r)}_R$ and $\psi^{(\bar{r})}_R$.
Besides the spinors $\varphi$ in $M^4$,
we should also introduce spinors $\chi$ on which the final factor acts.
We then obtain the following possible Weyl spinors:
\beqa
\varphi_R \otimes \psi^{(r)}_R \otimes \chi_R \ , \ \
\varphi_L \otimes \psi^{(\bar{r})}_R \otimes \chi_L \ , \label{RRrLLbr}\\
\varphi_L \otimes \psi^{(r)}_R \otimes \chi_L \ , \ \
\varphi_R \otimes \psi^{(\bar{r})}_R \otimes \chi_R \ , \label{LLrRRbr}\\
\varphi_R \otimes \psi^{(r)}_R \otimes \chi_L \ , \ \
\varphi_L \otimes \psi^{(\bar{r})}_R \otimes \chi_R \ , \label{RLrLRbr}\\
\varphi_L \otimes \psi^{(r)}_R \otimes \chi_R \ , \ \
\varphi_R \otimes \psi^{(\bar{r})}_R \otimes \chi_L \ . \label{LRrRLbr}
\eeqa
The two spinors in (\ref{RRrLLbr}) are in the charge conjugate representations
to each other.
So are those in (\ref{LLrRRbr}), (\ref{RLrLRbr}), and (\ref{LRrRLbr}).
We show it in detail in appendix~\ref{sec:chargecon}.

Since the IIB matrix model has the ten-dimensional Majorana-Weyl spinor,
we now impose these conditions.
By the Weyl condition, (\ref{RRrLLbr}) and (\ref{LLrRRbr}),
or (\ref{RLrLRbr}) and (\ref{LRrRLbr}), are chosen.
By the Majorana condition, the two spinors in 
(\ref{RRrLLbr})--(\ref{LRrRLbr})
are identified.
We still have two spinors, however.
We then obtain nonchiral spectrum.

There are two reasons why we could not obtain chiral spectrum.
First,
since we now consider four-dimensional topological configurations,
the zero modes of the same chirality,
$\psi^{(r)}_R$ and $\psi^{(\bar{r})}_R$, are obtained.
As the case $M^4 \times X^4 \subset M^8$
gave nonchiral spectrum in $M^4$,
now the first spinor in (\ref{RRrLLbr})
and the second spinor in (\ref{LLrRRbr}) 
necessarily arise and give nonchiral spectrum.

Second,
the {\it remainder} two dimensions $M^{10}/(M^4 \times S^2 \times S^2)$ interrupt.
In the gamma matrices (\ref{gammaM4S2S2}),
the ten-dimensional chirality operator becomes
\beq
\Gamma_{11} = \gamma_5 \otimes \id_2 \otimes \id_2 \otimes \sigma_3 \ .
\label{Gamma11}
\eeq
Then, even if $\Gamma_{11}=+$ is imposed, 
both $(\gamma_5, \sigma_3)=(+,+)$ and $(\gamma_5, \sigma_3)=(-,-)$ are allowed.
For instance, the first spinor in (\ref{RRrLLbr}) and the first spinor in (\ref{LLrRRbr}) appear.

Actually, the chirality on $S^2 \times S^2$, {\it i.e.}, whether 
one takes $\psi^{(r)}_R$ and $\psi^{(\bar{r})}_R$ or $\psi^{(r)}_L$ and $\psi^{(\bar{r})}_L$,
gives no difference.
Moreover, the chirality on each $S^2$ is irrelevant.
While the chirality operator on $S^2$ is $\gamma = n\cdot\sigma$,
the gamma matrix in the direction normal to $S^2$ is also
$\gamma_\bot = n\cdot\sigma$,
and their product gives $\gamma \gamma_\bot = \id_2$
in (\ref{Gamma11}).
Then, even if one considers a chiral mode on $S^2$, 
either $\gamma \psi =+\psi$ or  $\gamma \psi =-\psi$,
it gives no effect on (\ref{Gamma11}).

\section{Conclusions and Discussions}
\label{sec:conclusion}
\setcounter{equation}{0}

In this paper, 
we provided the GW Dirac operators and the index theorems on the fuzzy 2-sphere for matter
in the adjoint representation of the gauge group.
We extended our formulation to topologically nontrivial configurations, such as
the TP monopoles, the general configurations with $U(2)/U(1)^2$,
and the configurations with $U(\sum_p k_p)/\prod_p U(k_p)$.
We can also extend it to fuzzy $S^2 \times S^2$, $S^2 \times S^2 \times S^2$, and so on. 
The topological charge defined on fuzzy $(S^2)^l$ in this way
gives us a noncommutative generalization of the $l$th Chern character
on $(S^2)^l$,
as was shown in \cite{Aoki:2009cv} for the fundamental matter.
We will report on it in a separate paper.

We then studied the embeddings of topological configurations 
in higher dimensional matrix models, such as the IIB matrix model,
and discussed whether chiral spectrum is really obtained in  our spacetime.
The formulations using the GW relation gave a firm foundation to
such studies.
The GW relation indeed ensures the existence of chiral zero modes
against any variations 
since the index is a topological quantity.
As a practical advantage,
we can calculate exact chiral zero modes,
not approximate ones.   
Unfortunately, however, we could not obtain chiral spectrum 
by the $M^4 \times S^2 \times S^2$ embeddings in the IIB matrix model.
We now discuss how to resolve this problem.

One may consider decoupling dynamically one of the fields 
$\varphi_R \otimes \psi^{(r)}_R \otimes \chi_R$ and  
$\varphi_R \otimes \psi^{(\bar{r})}_R \otimes \chi_R$.
(See, for instance, ref. \cite{Poppitz:2010at}.)
By introducing strong coupling interactions, such as four-Fermi interactions,
to only one of them,
confinement may take place,
which makes all the composites massive and decoupled.
The other partner remains chiral and massless.
However,
introducing those interactions
seems artificial and unnatural
from the viewpoint that we derive everything from the IIB matrix model,
though it is allowed for formulating chiral gauge theories
on the lattice as in \cite{Poppitz:2010at}.

A simple way to obtain chiral spectrum in our spacetime is
to consider topological 
configurations in the entire extra six dimensions,
as we studied $M^4 \times X^6 \subset M^{10}$ 
in section \ref{sec:embedding}.
Coset space constructions, 
which cause
the ``remainder" dimensions,
are not suitable for it.
Torus  
is possible to construct 
in the same way as we did in this paper\footnote{The GW 
relation was implemented on the noncommutative 
torus by using the Neuberger's overlap Dirac 
operator \cite{Nishimura:2001dq}.  
In \cite{Iso:2002jc},
this GW Dirac operator was obtained from the general prescription 
of \cite{AIN2} and analyzed.
In \cite{Aoki:2008ik}, 
it was extended to the gauge fields in topologically
nontrivial sectors.
Dynamics of topological aspects in gauge theory on the noncommutative torus
were studied in \cite{Aoki:2006sb}.}.
Six-dimensional curved spaces can be described within 
six matrices in the formulation given in \cite{Hanada:2005vr}.
One may also consider situations similar to the
intersecting D-branes \cite{Ibanez:2001nd}, where 
one has no remainder dimensions normal to all of the D-branes 
which are intersecting to one another.
By T-duality, those situations are essentially equivalent to
the above ones.
We can also consider orbifolds in six dimensions
\cite{Aoki:2002jt,Chatzistavrakidis:2010xi}.
Imposing orbifold conditions plays the same role as 
the topological configurations giving the index.
We will report on these studies in future publications.

While we assumed the specific backgrounds in this paper,
we can in principle analyze whether such configurations are realized dynamically,
as we did in the analyses for the spacetime structures in the IIB matrix model
and in the analyses for the fuzzy spheres.
From such studies,
we might be able to find that
the standard model or its extension is obtained as a unique solution from
the IIB matrix model or its variants.
Or,
more complicated structures of the vacuum, such as the landscape, 
might be found,
but with the definite measure which enables us to 
discuss entropy.
Anyway, the matrix models make these studies possible.

\section*{Acknowledgements}
The author would like to thank Satoshi Iso and Jun Nishimura for discussions.

\appendix

\section{Commutative limits of Dirac operator and topological charge}
\label{sec:comlim_D_TC}
\setcounter{equation}{0}

In this appendix, we take
the commutative limits of the Dirac operator and the topological charge, 
and provide (\ref{DGWcom}) and (\ref{com_top_char_trivial}).
While similar calculations were given in \cite{AIN2,AIN3} for $\CTr[\hat\Gamma]$,
a coefficient becomes slightly different in this case,
and the calculation of $\CTr[\hat\Gamma]$ is also instructive for that of $\CTr[\Gamma]$.
We then show both calculations in a self-contained manner.

By substituting (\ref{defcovder}) into (\ref{def_hatG}),
we obtain
\beq
H_l = \sigma \cdot L^L + \frac{1}{2} +\rho \sigma \cdot a^L \ ,
\eeq
\beq
(H_l)^2 = \frac{n^2}{4} 
+\rho\Bigl(\{L_i^L,a_i^L\} + i \epsilon_{ijk}\sigma_k[L_i^L,a_j^L]+\sigma \cdot a^L \Bigr)
+\rho^2(\sigma \cdot a^L)^2 \ ,
\eeq
and
\beqa
\hat\Gamma &= &a(\sigma \cdot L^L + \frac{1}{2} +\rho \sigma \cdot a^L) 
-\frac{1}{2}a^3 \rho \sigma \cdot L^L \{L_i^L, a_i^L \} \n
&&-\frac{1}{2}a^3\rho \sigma \cdot L^L \left( i \epsilon_{ijk}\sigma_k[L_i^L,a_j^L]
+\sigma \cdot a^L+\rho(\sigma \cdot a^L)^2 
-\frac{3}{4}a^2\rho \{L_i^L, a_i^L \}^2
\right) \n
&&-\frac{1}{2}a^3\rho (\frac{1}{2}+\rho \sigma \cdot a^L) \{L_i^L, a_i^L \} \n
&&+{\cal O}(n^{-3}) \ ,
\label{hatG_expand_n}
\eeqa
with $a=\frac{2}{n}$.
Similarly, by substituting (\ref{defcovder}) into (\ref{def_G}),
we obtain 
\beqa
\Gamma &= &a(\sigma \cdot L^R - \frac{1}{2} +\rho \sigma \cdot a^R) 
-\frac{1}{2}a^3 \rho \sigma \cdot L^R \{L_i^R, a_i^R \} \n
&&-\frac{1}{2}a^3\rho \sigma \cdot L^R \left( 
 i \epsilon_{ijk}\sigma_k[L_i^R,a_j^R]
-\sigma \cdot a^R +\rho(\sigma \cdot a^R)^2
-\frac{3}{4}a^2\rho \{L_i^R, a_i^R \}^2
\right) \n
&&-\frac{1}{2}a^3\rho (-\frac{1}{2}+\rho \sigma \cdot a^R) \{L_i^R, a_i^R \} \n
&&+{\cal O}(n^{-3}) \ ,
\label{G_expand_n}
\eeqa
For the commutative limit of the Dirac operator (\ref{defDGW}),
it is enough to take terms up to order $n^{-1}$ in 
(\ref{hatG_expand_n}) and (\ref{G_expand_n}).
We then easily obtain (\ref{DGWcom}).

For the commutative limit of the topological charge, the rhs of 
(\ref{ITtrivial}), however,
we should take terms up to order $n^{-2}$ in 
(\ref{hatG_expand_n}) and (\ref{G_expand_n}),
since $\CTr$ gives a contribution of order $n^2$.
We first consider $\CTr[\hat\Gamma]$.
Taking the trace over the spinor index, we obtain
\beq
\CTr[\hat\Gamma]=
\CTr' \left[ \frac{2}{n}-a^3\rho\left(L^L_k i\epsilon_{ijk}[L^L_i, a^L_j]+L^L_i a^L_i
+i\rho\epsilon_{ijk}L_i^L a_j^L a_k^L
+\frac{1}{2} \{L^L_i, a^L_i \} \right)\right] \ ,
\eeq
where $\CTr'$ is the trace over the whole configuration space without 
the spinor index.
It is rewritten as 
$\CTr'=\tr_L \tr_{t_L} \tr_R \tr_{t_R}$,
where $\tr_L$ is the trace over the space on which $L_i^L$ act,
$\tr_{t_L}$ is the trace over the space on which 
the gauge group generators $(t^a)^L$ act,
and so on.
In the commutative limit,
$\frac{1}{n}\tr_L(M^L)$ is replaced by $\int\frac{d\Omega_L}{4\pi} M(\Omega_L)$,
and $\frac{1}{n}\tr_R(M^R)$ by $\int\frac{d\Omega_R}{4\pi} M(\Omega_R)$.
Then, $\CTr'$ becomes
$n^2\int\frac{d\Omega_L}{4\pi}\int\frac{d\Omega_R}{4\pi} \tr_{t_L} \tr_{t_R}$.
It then follows that
\beqa
\CTr[\hat\Gamma] &\to&
\int\frac{d\Omega_L}{4\pi}\int\frac{d\Omega_R}{4\pi} \tr_{t_L} \tr_{t_R}
\left(2n + 2\rho^2\epsilon_{ijk} n^L_i F^L_{jk}\right) \n
&=&2nk^2 + 2\rho^2 \int\frac{d\Omega}{4\pi} \tr
\left(\epsilon_{ijk} n_i F_{jk}\right) \ .
\label{comlim_hatG}
\eeqa
where
$F_{ij}= \partial_i a_j'-\partial_j a_i'-i[a_i',a_j']$
with $a'_i$ 
given in (\ref{decomposeto}).
In the last line, we used a simple expression $\tr = \tr_{t_L} \tr_{t_R}$.

Similarly, we obtain
\beq
\CTr[\Gamma]=
\CTr' \left[ -\frac{2}{n}-a^3\rho\left(L^R_k i\epsilon_{ijk}[L^R_i, a^R_j]-L^R_i a^R_i
+i\rho\epsilon_{ijk}L_i^R a_j^R a_k^R
-\frac{1}{2} \{L^R_i, a^R_i \} \right)\right] \ ,
\eeq
and then
\beqa
\CTr[\Gamma] &\to&
\int\frac{d\Omega_L}{4\pi}\int\frac{d\Omega_R}{4\pi} \tr_{t_L} \tr_{t_R}
\left(-2n - 2\rho^2\epsilon_{ijk} n^R_i F^R_{jk}\right) \n
&=&-2nk^2 - 2\rho^2 \int\frac{d\Omega}{4\pi} \tr
\left(\epsilon_{ijk} n_i F_{jk}\right) \ .
\label{comlim_G}
\eeqa
Because of the relation $[A^R, B^R] = -[A, B]^R$,
there arose the minus sign in front of the field strength $F_{jk}$ in (\ref{comlim_G}),
compared with (\ref{comlim_hatG}).
Adding (\ref{comlim_hatG}) and (\ref{comlim_G}),
we finally obtain (\ref{com_top_char_trivial}).

\section{General configurations with $U(\sum_p k_p)/\prod_p U(k_p)$}
\label{sec:genecon_Uk}
\setcounter{equation}{0}

In this appendix, we study formulations for general configurations
with $U(\sum_p k_p)/\prod_p U(k_p)$.
In particular, we show that
the commutative limits of the GW Dirac operators and the topological charges
have similar forms as 
(\ref{com_lim_Dpr})--(\ref{comlim_TCgenl}).

As we mentioned at the end of section \ref{sec:config_k},
for each electric charge operator $T'_p$  with $p=1, \ldots, h-1$, given by (\ref{defTgenka}),
we define modified chirality operators 
$\Gamma'_{pr}$, $\hat\Gamma'_{pr}$, $\Gamma'_{pl}$ and $\hat\Gamma'_{pl}$
by (\ref{def_Gr})--(\ref{def_hatGl}).
We then define modified GW Dirac operators
$D'_{pr}$ and $D'_{pl}$
by (\ref{def_DGWgenr}) and (\ref{def_DGWgenl}).
They satisfy the GW relations as (\ref{GWrel_Dpr}) and (\ref{GWrel_Dpl}), 
and the index theorems as (\ref{ITgenr}) and (\ref{ITgenl}).

We now study the commutative limits.
Following (\ref{comlimTprime}), 
we write the commutative limits of 
the electric charge operators $T'_p$ as
\beq
T'_p \to 2 \phi'_p=\sum_a 2\phi'^a_p t^a \ ,
\eeq
where $t^a$ are the generators of the gauge group $U(\sum_{p=1}^h k_p)$.
Because of $(T'_p)^2=1$, 
\beq
\sum_{a,b} \phi'^a_p \phi'^b_p t^a t^b =\frac{1}{4} 
\eeq
should be satisfied at the commutative level as well.
The rhs is the identity operator in the gauge group space
and the coordinate space of the sphere.
Then, unlike the $U(2)$ case,
$\phi'^a_p = (1,0,\ldots,0)$ gauge does not exist in general,
though we have gauges where all of  $\phi'^a_p$ are constant and
independent of the sphere coordinate $\Omega$.

The commutative limit of the GW Dirac operator $D'_{pr}$ becomes
\beq
D'_{pr} \ \to \frac{1}{2} \{2\phi'^R_p , (\sigma_i {\cal L}_i +1) \}
+ \frac{1}{2} \{2\phi'^R_p  , \rho \sigma_i P_{ij} a_j^L \} 
-\frac{1}{2} \{2\phi'^R_p  , \rho \sigma_i P_{ij} a_j^R \}  \ ,
\label{com_lim_Dppr}
\eeq
as (\ref{com_lim_Dpr}).
The superscript $R$ ($L$) means that the operator acts 
from right (left) in the gauge group space:
$\phi'^R_p = \phi'^a_p(\Omega)(t^a)^R$, {\it etc}.
In the gauges $\phi'^a_p(\Omega) = \phi'^a_p$,
where $\phi'^a_p$ are constant,
(\ref{com_lim_Dppr}) becomes
\beq
2\phi'^R_p (\sigma_i {\cal L}_i +1
+ \rho \sigma_i P_{ij} a_j^L)
-\phi'^a_p \rho \sigma_i P_{ij} a_j^b \{t^a,t^b\}^R
\equiv D'_{pr, {\rm com}} \ .
\eeq
This Dirac operator has the adjoint coupling of the
unbroken $U(1)$ gauge field $\sum_a \phi'^a_p a_j^a (t^a)^R \tilde{(t^a)}$.
It also satisfies a chiral relation 
\beq
\{D'_{pr,{\rm com}}  ,  \gamma'_{pr} \} = 0 \ ,
\eeq
where $ \gamma'_{pr} = 2\phi'^R_p \gamma$ is the chirality operator
multiplied by the unbroken $U(1)$ charge.
The same arguments hold also for $D'_{pl}$.

As (\ref{comlim_TCgenr}) and (\ref{comlim_TCgenl}),
the commutative limits of the topological charges become
\beqa
\frac{1}{2} {\cal T}r [\Gamma'_{pr} + \hat\Gamma'_{pr}]
\to
-2k\frac{\rho^2}{8\pi}\int d\Omega \epsilon_{ijk}
n_i \Bigl( \phi'^a_p F_{jk}^a 
- f_{abc} \phi'^a_p (D_j \phi'_p)^b (D_k \phi'_p)^c \Bigr) \ ,
\label{comlim_TCgenpr} \\
\frac{1}{2} {\cal T}r [\Gamma'_{pl} + \hat\Gamma'_{pl}]
\to
2k\frac{\rho^2}{8\pi}\int d\Omega \epsilon_{ijk}
n_i \Bigl( \phi'^a_p F_{jk}^a 
- f_{abc} \phi'^a_p (D_j \phi'_p)^b (D_k \phi'_p)^c \Bigr) \ ,
\label{comlim_TCgenpl} 
\eeqa
where $k=\sum_{p=1}^h k_p$ 
and $f_{abc}$ are the structure constants of $U(\sum_{p=1}^h k_p)$.
The field strength $F_{jk}=F_{jk}^a t^a$ 
is defined as
$F_{jk}= \partial_j a_k'-\partial_k a_j'-i[a_j',a_k']$,
and the covariant derivative $D_{j}$ 
is defined as
$D_{j} =\partial_{j}  -i[a'_{j} , ~~]$, with
$a'_j$ given in (\ref{decomposeto}).
In the gauges $\phi'^a_p(\Omega) = \phi'^a_p$,
where $\phi'^a_p$ are constant,
the integrand of 
(\ref{comlim_TCgenpr}) and (\ref{comlim_TCgenpl}) 
indeed gives the Abelian flux in the unbroken $U(1)$ direction
$\phi'^a_p(\partial_j a'^a_k - \partial_k a'^a_j)$.

We finally give a comment.
We here obtained the $h-1$ topological charges 
${\cal T}r [\Gamma'_{pr} + \hat\Gamma'_{pr}]$
 with $1 \le p \le h-1$,
while we had $\frac{h(h-1)}{2}$ ones (\ref{ITconfka})
for $1 \le p <q \le h$.
The lack of information is covered by defining
chirality operators
\beqa
\Gamma'_{p,q} &=&T'^L_p 
\frac{\{T'^R_q , \Gamma\}} {\sqrt{\{T'^R_q , \Gamma\}^2}} \ ,  \\
\hat\Gamma'_{p,q} &=&
\frac{\{T'^L_p , \hat\Gamma\}} {\sqrt{\{T'^L_p , \hat\Gamma\}^2}} T'^R_q \ ,
\eeqa
and GW Dirac operators
\beq
D'_{p,q} = -a^{-1} \Gamma'_{p,q}(1-\Gamma'_{p,q} \hat\Gamma'_{p,q}) \ ,
\eeq
for $1 \le p,q \le h-1$.
They satisfy GW relations and then index theorems
\beq
{\rm index}(D'_{p,q}) =
\frac{1}{2} {\cal T}r [\Gamma'_{p,q} + \hat\Gamma'_{p,q}] \ ,
\eeq
which indeed provide $\frac{(h-1)(h-2)}{2}$ topological charges.
While ${\cal T}r (\Gamma'_{p,q})$ and ${\cal T}r (\hat\Gamma'_{p,q})$
vanish for the $U(2)/U(1)^2$ case of section \ref{sec:genecon211},
they give nontrivial results in the present case
of $U(\sum_p k_p)/\prod_p U(k_p)$.

\section{Charge conjugation}
\label{sec:chargecon}
\setcounter{equation}{0}

In this appendix we show that the two spinors in
(\ref{RRrLLbr})--(\ref{LRrRLbr})
are in the charge conjugate representations to each other.
We also show that the Majorana condition in ten dimensions
can be written as the decomposition into each subspace,
as in the Weyl condition.

We first introduce unitary matrices $B_1$ and $B_2$ 
acting on $SO(9,1)$ spinors, which satisfy
\beqa
B_1 \Gamma_M B_1^{-1} &=& (\Gamma_M)^* \ , 
\label{def_B1} \\
B_2 \Gamma_M B_2^{-1} &=& -(\Gamma_M)^* \ ,
\label{def_B2}
\eeqa
for $M=0,\ldots, 9$.
(We follow the notation in Appendix B.1 in \cite{Polchinski:1998rr}.)
For the representation of gamma matrices (\ref{gammaM4S2S2}),
they are written as
\beqa
B_1 &=& B_1^{(4)} \otimes \sigma_2 \otimes \sigma_2 \otimes \sigma_2 \ , 
\label{B1_explicit_form} \\
B_2 &=& B_2^{(4)} \otimes \sigma_2 \otimes \sigma_2 \otimes \sigma_1 \ ,
\label{B2_explicit_form}
\eeqa
where $B_1^{(4)}$ and $B_2^{(4)}$ satisfy
\beqa
B_1^{(4)} \gamma_\mu (B_1^{(4)})^{-1} &=& -(\gamma_\mu)^* \ ,\\
B_2^{(4)} \gamma_\mu (B_2^{(4)})^{-1} &=& (\gamma_\mu)^* \ .
\eeqa
The charge conjugation of $SO(9,1)$ spinors is defined as
\beq
\zeta^C \equiv B^{-1} \zeta^* \ ,
\eeq
for either $B=B_1$ or $B=B_2$.

For the gamma matrices (\ref{gammaM4S2S2}),
the chirality operator in $M^{10}$ is written as
\beq
\Gamma_{11} = \gamma_5 \otimes \id_2 \otimes \id_2 \otimes \sigma_3 \ ,
\label{Gamma11_deco}
\eeq
where the chirality operator in $M^4$ is 
\beq
\gamma_5 = -i \gamma_0 \gamma_1 \gamma_2 \gamma_3 \ .
\eeq
As usual,
\beq
B \Gamma_{11} B^{-1} = (\Gamma_{11})^*
\eeq
is satisfied for both $B_1$ and $B_2$, 
while
\beq
B^{(4)} \gamma_5 (B^{(4)})^{-1} = -(\gamma_5)^*
\label{gamma5_complex}
\eeq
is satisfied for both $B_1^{(4)}$ and $B_2^{(4)}$.
Then, the Weyl spinor in $M^{10}$ is self-conjugate
and that in $M^4$ is complex.

We may define
a chirality operator
in the second and the third factors in (\ref{Gamma11_deco}) as
\beq
\Gamma^{(R^3 \times R^3)}= \id_2 \otimes \id_2 \ .
\label{def_Gamma_6} 
\eeq
We can also define chirality operators in this space as
\beqa
\Gamma^{(S^2\times S^2)}&=& n \cdot \sigma \otimes  n \cdot \sigma \ , 
\label{def_Gamma_S2S2} \\
\Gamma^{(S^2)}&=&  n \cdot \sigma \otimes \id_2 \ ,
\label{def_Gamma_S2} \\
\Gamma^{(S^{2 \prime})}&=& \id_2 \otimes n \cdot \sigma \ .
\label{def_Gamma_S2_p}
\eeqa
The charge conjugation matrix in this space is
\beq
A=\sigma_2 \otimes \sigma_2
\eeq
for either (\ref{B1_explicit_form}) or (\ref{B2_explicit_form}).
The Weyl spinor in terms of the chirality (\ref{def_Gamma_6}) 
is self-conjugate because 
\beq
A \Gamma^{(R^3 \times R^3)} A^{-1} = (\Gamma^{(R^3 \times R^3)})^* \ 
\eeq
is satisfied.
That of (\ref{def_Gamma_S2S2}) is self-conjugate:
\beq
A \Gamma^{(S^2\times S^2)} A^{-1} = (\Gamma^{(S^2\times S^2)})^* \ ,
\label{S2S2_self_conj}
\eeq
and those of (\ref{def_Gamma_S2}) and (\ref{def_Gamma_S2_p}) are complex:
\beq
A \Gamma^{(S^2)} A^{-1} = -(\Gamma^{(S^2)})^* \ .
\eeq

We should also define a chirality operator in 
the fourth factor in (\ref{Gamma11_deco}) as
\beq
\Gamma^{(e)} = \sigma_3 \ .
\eeq
The charge conjugation matrix in this space is
\beq
A^{(e)}_1 = \sigma_2 \ , \ \
A^{(e)}_2 = \sigma_1
\eeq
for (\ref{B1_explicit_form}) and (\ref{B2_explicit_form}), respectively.
For either $A^{(e)}_1$ or $A^{(e)}_2$,
the Weyl spinor is complex because 
\beq
A^{(e)} \Gamma^{(e)} (A^{(e)})^{-1} = -(\Gamma^{(e)})^* \ .
\label{Gamma_e_complex}
\eeq
It follows from (\ref{gamma5_complex}), 
(\ref{S2S2_self_conj}) and (\ref{Gamma_e_complex})
that the two spinors in
(\ref{RRrLLbr})--(\ref{LRrRLbr})
are in the charge conjugate representations to each other.

In the remainder of this appendix,
we discuss the Majorana condition.
The Majorana condition in ten dimensions 
\beq
\zeta = \zeta^C \equiv B^{-1} \zeta^* 
\label{maj_con_M10}
\eeq
can be imposed since $B^* B=1$ is satisfied for
either $B=B_1$ in (\ref{def_B1})
or $B=B_2$ in (\ref{def_B2}).

By decomposing the spinor as
\beq
\zeta = \varphi \otimes \psi \otimes \chi \ ,
\eeq
the Majorana condition (\ref{maj_con_M10}) with $B_2$ in (\ref{B2_explicit_form})
is written as
\beq
\varphi^* \otimes \psi^* \otimes \chi^* 
=B_2^{(4)}\varphi \otimes A \psi \otimes A^{(e)}_2 \chi \ .
\eeq
This is satisfied by imposing the conditions
\beqa
\varphi^* &=& \pm B_2^{(4)}\varphi \ , 
\label{B2_dec_con_1}\\
\psi^* &=&  \pm A \psi \ , 
\label{B2_dec_con_23} \\
\chi^* &=& \pm A^{(e)}_2 \chi \ ,
\label{B2_dec_con_4}
\eeqa
where the three signs should satisfy $(\pm)(\pm)(\pm)=+$.
Since $(B_2^{(4)})^* B_2^{(4)}=1$ and $(A^{(e)}_2)^* A^{(e)}_2=1$
are satisfied, 
(\ref{B2_dec_con_1}) and (\ref{B2_dec_con_4}) can be imposed.
While
the reality condition,
the Euclidean version of the Majorana condition, can not be imposed on the $SO(3)$ spinors,
which are in the pseudoreal representation,
the product of two pseudoreal representations is real.
This trick is used in (\ref{B2_dec_con_23}), where $A^* A =1$ is satisfied.

Similarly, 
the Majorana condition (\ref{maj_con_M10}) with $B_1$ in (\ref{B1_explicit_form})
is written as
\beq
\varphi^* \otimes \psi^* \otimes \chi^* 
=B_1^{(4)}\varphi \otimes A \psi \otimes  A^{(e)}_1 \chi \ .
\eeq
This is satisfied by imposing the conditions
\beqa
\varphi^* \otimes \chi^* 
&=& \pm B_1^{(4)}\varphi \otimes  A^{(e)}_1 \chi \ ,
\label{B1_dec_con_14} \\
\psi^* &=& \pm A \psi \ , 
\label{B1_dec_con_23} 
\eeqa
where the two signs should satisfy $(\pm)(\pm)=+$.
The trick of doubling the pseudoreal representations
is used twice, in (\ref{B1_dec_con_14}) and in (\ref{B1_dec_con_23}).

We therefore find that the Majorana condition in ten dimensions
can be written as the decomposition into each subspace:
(\ref{B2_dec_con_1})--(\ref{B2_dec_con_4}),
or (\ref{B1_dec_con_14}) and (\ref{B1_dec_con_23}).
Although these decompositions were not used directly in the present paper,
they are useful when we study the Majorana condition in each subspace.


\begin{thebibliography}{99}
\bibitem{Banks:1996vh}
T.~Banks, W.~Fischler, S.~H.~Shenker and L.~Susskind,
Phys.\ Rev.\ D {\bf 55}, 5112 (1997)
[arXiv:hep-th/9610043].

\bibitem{IKKT}
N.~Ishibashi, H.~Kawai, Y.~Kitazawa and A.~Tsuchiya,
Nucl.\ Phys.\ B {\bf 498}, 467 (1997)
[arXiv:hep-th/9612115].
%
For a review:
H.~Aoki, S.~Iso, H.~Kawai, Y.~Kitazawa, A.~Tsuchiya and T.~Tada,
Prog.\ Theor.\ Phys.\ Suppl.\  {\bf 134}, 47 (1999)
[arXiv:hep-th/9908038].

\bibitem{Aoki:1998vn}
  H.~Aoki, S.~Iso, H.~Kawai, Y.~Kitazawa and T.~Tada,
  Prog.\ Theor.\ Phys.\  {\bf 99}, 713 (1998)
  [arXiv:hep-th/9802085].

\bibitem{Nishimura:2001sx}
  J.~Nishimura and F.~Sugino,
  JHEP {\bf 0205}, 001 (2002)
  [arXiv:hep-th/0111102].

\bibitem{AIMN}
H.~Aoki, S.~Iso, T.~Maeda and K.~Nagao,
  Phys.\ Rev.\ D {\bf 71}, 045017 (2005)
  [arXiv:hep-th/0412052]

\bibitem{Steinacker:2007ay}
  P.~Aschieri, T.~Grammatikopoulos, H.~Steinacker and G.~Zoupanos,
  JHEP {\bf 0609}, 026 (2006)
  [arXiv:hep-th/0606021];
%
  H.~Steinacker and G.~Zoupanos,
  JHEP {\bf 0709}, 017 (2007)
  [arXiv:0706.0398 [hep-th]];

\bibitem{Chatzistavrakidis:2009ix}
  A.~Chatzistavrakidis, H.~Steinacker and G.~Zoupanos,
  Fortsch.\ Phys.\  {\bf 58}, 537 (2010)
  [arXiv:0909.5559 [hep-th]].

\bibitem{Atiyah:1971rm}
  M.~F.~Atiyah and I.~M.~Singer,
  Annals Math.\  {\bf 93}, 139 (1971).

\bibitem{Nielsen:1980rz}
  H.~B.~Nielsen and M.~Ninomiya,
  Nucl.\ Phys.\  B {\bf 185}, 20 (1981)
  [Erratum-ibid.\  B {\bf 195}, 541 (1982)];
%
  Nucl.\ Phys.\  B {\bf 193}, 173 (1981).

\bibitem{GinspargWilson}
P.~H.~Ginsparg and K.~G.~Wilson,
Phys.\ Rev.\ D {\bf 25}, 2649 (1982).

H.~Neuberger,
Phys.\ Lett.\ B {\bf 417}, 141 (1998)
[arXiv:hep-lat/9707022];
%
Phys.\ Rev.\ D {\bf 57}, 5417 (1998)
[arXiv:hep-lat/9710089];
%
Phys.\ Lett.\ B {\bf 427}, 353 (1998)
[arXiv:hep-lat/9801031].

M.~L\"uscher,
Phys.\ Lett.\ B {\bf 428}, 342 (1998)
[arXiv:hep-lat/9802011].

P.~Hasenfratz,
Nucl.\ Phys.\ Proc.\ Suppl.\  {\bf 63}, 53 (1998)
[arXiv:hep-lat/9709110];
%
P.~Hasenfratz, V.~Laliena and F.~Niedermayer,
Phys.\ Lett.\ B {\bf 427}, 125 (1998)
[arXiv:hep-lat/9801021];
%
F.~Niedermayer,
Nucl.\ Phys.\ Proc.\ Suppl.\  {\bf 73}, 105 (1999)
[arXiv:hep-lat/9810026].

\bibitem{Luscher:1981zq}
  M.~L\"uscher,
  Commun.\ Math.\ Phys.\  {\bf 85}, 39 (1982);
%
  Nucl.\ Phys.\  B {\bf 549}, 295 (1999)
  [arXiv:hep-lat/9811032].

\bibitem{AIN2}
H.~Aoki, S.~Iso and K.~Nagao,
Phys.\ Rev.\ D {\bf 67}, 085005 (2003)
[arXiv:hep-th/0209223].
%

\bibitem{Madore}
J.~Madore,
Class.\ Quant.\ Grav.\  {\bf 9}, 69 (1992).
   
\bibitem{balagovi}
A.~P.~Balachandran, T.~R.~Govindarajan and B.~Ydri,
Mod.\ Phys.\ Lett.\ A {\bf 15}, 1279 (2000)
[arXiv:hep-th/9911087];
%
arXiv:hep-th/0006216.

\bibitem{Balachandran:2003ay}
A.~P.~Balachandran and G.~Immirzi,
Phys.\ Rev.\ D {\bf 68}, 065023 (2003)
[arXiv:hep-th/0301242].

\bibitem{AIN3}
H.~Aoki, S.~Iso and K.~Nagao,
Nucl.\ Phys.\ B {\bf 684}, 162 (2004)
[arXiv:hep-th/0312199].
%

\bibitem{AIM}
  H.~Aoki, S.~Iso and T.~Maeda,
  Phys.\ Rev.\  D {\bf 75}, 085021 (2007)
  [arXiv:hep-th/0610125].

\bibitem{Aoki:2008qta}
  H.~Aoki, Y.~Hirayama and S.~Iso,
  Phys.\ Rev.\  D {\bf 78}, 025028 (2008)
  [arXiv:0804.0568 [hep-th]].
For a review:
  H.~Aoki,
  Prog. \ Theor. \ Phys. \ Suppl. \ {\bf 171}, 228 (2007)
  [arXiv:0706.3078 [hep-th]].
  
 \bibitem{'tHooft:1974qc}
  G.~'t Hooft,
  Nucl.\ Phys.\ B {\bf 79}, 276 (1974).
 
\bibitem{Grosse:2010zq}
  H.~Grosse, F.~Lizzi and H.~Steinacker,
  Phys.\ Rev.\  D {\bf 81}, 085034 (2010)
  [arXiv:1001.2703 [hep-th]].
 
\bibitem{ISTT}
  G.~Ishiki, S.~Shimasaki, Y.~Takayama and A.~Tsuchiya,
  JHEP {\bf 0611}, 089 (2006)
  [arXiv:hep-th/0610038];
%
  T.~Ishii, G.~Ishiki, S.~Shimasaki and A.~Tsuchiya,
  JHEP {\bf 0705}, 014 (2007)
  [arXiv:hep-th/0703021];
%
  Phys.\ Rev.\  D {\bf 77}, 126015 (2008)
  [arXiv:0802.2782 [hep-th]].

\bibitem{Aoki:2009cv}
  H.~Aoki, Y.~Hirayama and S.~Iso,
  Phys.\ Rev.\  D {\bf 80}, 125006 (2009)
  [arXiv:0909.5252 [hep-th]].

  
\bibitem{Poppitz:2010at}
  E.~Poppitz and Y.~Shang,
  Int.\ J.\ Mod.\ Phys.\  A {\bf 25}, 2761 (2010)
  [arXiv:1003.5896 [hep-lat]].
\bibitem{Nishimura:2001dq}
J.~Nishimura and M.~A.~Vazquez-Mozo,
JHEP {\bf 0108}, 033 (2001)
[arXiv:hep-th/0107110].

\bibitem{Iso:2002jc}
S.~Iso and K.~Nagao,
Prog.\ Theor.\ Phys.\  {\bf 109}, 1017 (2003)
[arXiv:hep-th/0212284].

\bibitem{Aoki:2008ik}
  H.~Aoki, J.~Nishimura and Y.~Susaki,
  JHEP {\bf 0904}, 055 (2009)
  [arXiv:0810.5234 [hep-th]].
  
\bibitem{Aoki:2006sb}
  H.~Aoki, J.~Nishimura and Y.~Susaki,
  JHEP {\bf 0702}, 033 (2007)
  [arXiv:hep-th/0602078];
%
  JHEP {\bf 0710}, 024 (2007)
  [arXiv:hep-th/0604093];
%
  JHEP {\bf 0909}, 084 (2009)
  [arXiv:0907.2107 [hep-th]].

\bibitem{Hanada:2005vr}
  M.~Hanada, H.~Kawai and Y.~Kimura,
  Prog.\ Theor.\ Phys.\  {\bf 114}, 1295 (2005)
  [arXiv:hep-th/0508211].
  
\bibitem{Ibanez:2001nd}
  M.~Berkooz, M.~R.~Douglas and R.~G.~Leigh,
  Nucl.\ Phys.\  B {\bf 480}, 265 (1996)
  [arXiv:hep-th/9606139];

  L.~E.~Ibanez, F.~Marchesano and R.~Rabadan,
  JHEP {\bf 0111}, 002 (2001)
  [arXiv:hep-th/0105155];

  R.~Blumenhagen, B.~Kors, D.~Lust and S.~Stieberger,
  Phys.\ Rept.\  {\bf 445}, 1 (2007)
  [arXiv:hep-th/0610327].
 
\bibitem{Aoki:2002jt}
  H.~Aoki, S.~Iso and T.~Suyama,
  Nucl.\ Phys.\  B {\bf 634}, 71 (2002)
  [arXiv:hep-th/0203277].

\bibitem{Chatzistavrakidis:2010xi}
  A.~Chatzistavrakidis, H.~Steinacker and G.~Zoupanos,
  JHEP {\bf 1005}, 100 (2010)
  [arXiv:1002.2606 [hep-th]].
  

\bibitem{Polchinski:1998rr}
  J.~Polchinski,
  ``String theory. Vol. 2: Superstring theory and beyond,''
{\it  Cambridge, UK: Univ. Pr. (1998) 531 p}

\end{thebibliography}
\end{document}